# Local Avalanche Photodetectors Driven by Lightning-rod Effect and Surface Plasmon Excitations

Zhao Fu[1,2], Meng Yuan[1], Jiafa Cai[1,3], Rongdun Hong[1,3], Xiaping Chen[1,3], Dingqu Lin[1,3], Shaoxiong Wu[1,3], Yuning Zhang[1], Zhengyun Wu[1,3], Zhanwei Shen[4]*, Zhijie Wang[4]*, Jicheng Wang[5]*, Mingkun Zhang[1,6]*, Zhilin Yang[1]*, Deyi Fu[1]*, Feng Zhang[1,3]*, Rong Zhang[1]*

zwshen@semi.ac.cn, wangzj@semi.ac.cn, jcwang@jiangnan.edu.cn, mkzhang@xmu.edu.cn,

zlyang@xmu.edu.cn, dyfu@xmu.edu.cn, fzhang@xmu.edu.cn, rzhangxmu@xmu.edu.cn

[1]Department of physics, Xiamen university, Fujian, 361005, P. R. China
[2]College of Electrical Engineering, Tongling university, Anhui, 244061, P. R. China
[3]Jiujiang Research Institute of Xiamen University, Jiangxi, 332000, P. R. China.
[4]Laboratory of Solid-State Optoelectronics Information Technology, Institute of Semiconductors, Chinese Academy of Sciences, Beijing, 100083, P. R. China
[5]School of Science, Jiangnan University, Jiangsu, 214122, China
[6]The Higher Educational Key Laboratory of Flexible Manufacturing Equipment Integration of Fujian Province, Xiamen Institute of Technology, Fujian, 361005, P. R. China

**Abstract:** Sensitive avalanche photodetectors (APDs) that operate within the ultraviolet spectrum are critically required for applications in detecting fire and deep-space exploration. However, the development of such devices faces significant challenges, including high avalanche breakdown voltage, the necessity for complex quenching circuits, and thermal runaway associated with Geiger-mode avalanche operation. To mitigate these issues, we report on a 4H-SiC APD design utilizing micro-holes (MHs) structures and Al nano-triangles (NTs) to enhance surface electric field driven by strong localized surface plasmon excitations and lightning-rod effect. The device demonstrates a record low avalanche breakdown voltage of approximately 14.5 V, a high detectivity of $7\times10^{13}$ Jones, a nanosecond-level response time, and repeated stable detections without the requirement of a quenching circuit. Collectively, when compared with the conventional wide-bandgap-based APDs, this device achieves a reduction in avalanche breakdown voltage by an order of magnitude and exhibits a substantial increase in detectivity. Consequently, the proposed APD configuration presents a promising candidate for ultraviolet detection and integrated optoelectronic circuits.

**Introduction**

Avalanche photodetectors (APDs) are widely utilized in various applications, including lidar, chemical sensing, flame detection, ozone-hole sensing, and telecommunications, owing to their intrinsic ultra-high gain, which enables high sensitivity with a range extending from ultraviolet to terahertz [1-13]. To date, ultraviolet APDs based on wide-bandgap materials as SiC[14,15], GaN[16-21], $Ga_2O_3$[22,23], and ZnO[24,25] have shown impressive performance, achieving high gain (>1000) and high response speed (<1 μs). These capabilities are largely attributed to their inherent properties, including high avalanche characteristics, low minority carrier lifetime, and intrinsic ultraviolet absorption[26]. Generally, conventional APDs typically operate in the Geiger mode under 95% of critical electrical field to achieve high gain. However, several critical limitations arise when operating APDs in this mode. First, the avalanche state increases the risk of device breakdown, necessitating the use of quenching circuits to prevent permanent damage[14,15,27-33]. This requirement increases considerable complexity to their practical application. Second, the long-term stability and operational life of traditional APDs are compromised due to the stress induced by continuous Geiger-mode operation[34]. Furthermore, wide-bandgap APDs functioning in Geiger mode require relatively high driving voltages (≥100 V), which makes them less suitable for applications that demand lower operational voltages. Therefore, it is critical to develop wide-bandgap APDs capable of achieving high gain at lower voltages for high stability, while eliminating the requirement for quenching circuits. Such advancements could significantly expand the range of APD applications.

An alternative approach on leveraging field enhancement is to construct semiconductor microstructures and metallic nanostructures, which have demonstrated considerable promise in enhancing local electric fields[35-40]. These enhancements arise from the accumulation of charges in the sharp zones of metallic nanostructures or near the corners of semiconductor microstructures, leading to localized amplification of electric field[41]. However, there are relatively few reports on combining the lightning-rod effect with surface plasmon excitations to enhance electric field to form an avalanche. By incorporating this field enhancement within semiconductor photodetectors, the internal electric field can reach avalanche field intensities (1–3 MV/cm) under lower reverse bias conditions.

In this study, a novel SiC-based APD design incorporating semiconductor microstructures and metallic nanostructures are presented and demonstrated with lightning-rod effect and surface plasmon excitation mechanisms. This design achieves a low breakdown voltage of 14.5 V with high detectivity of $7\times10^{13}$ Jones, high gain of $10^4$, and an improved response speed of nanoseconds. These advancements enable scalable and flexible platforms within mature SiC device technology, paving the way for the development of APDs capable of detecting weak signals without Geiger-mode operation. Moreover, the approach has broader implications, potentially influencing the design of wide-bandgap optoelectronic devices, quantum devices, and integrated optoelectronic circuits.

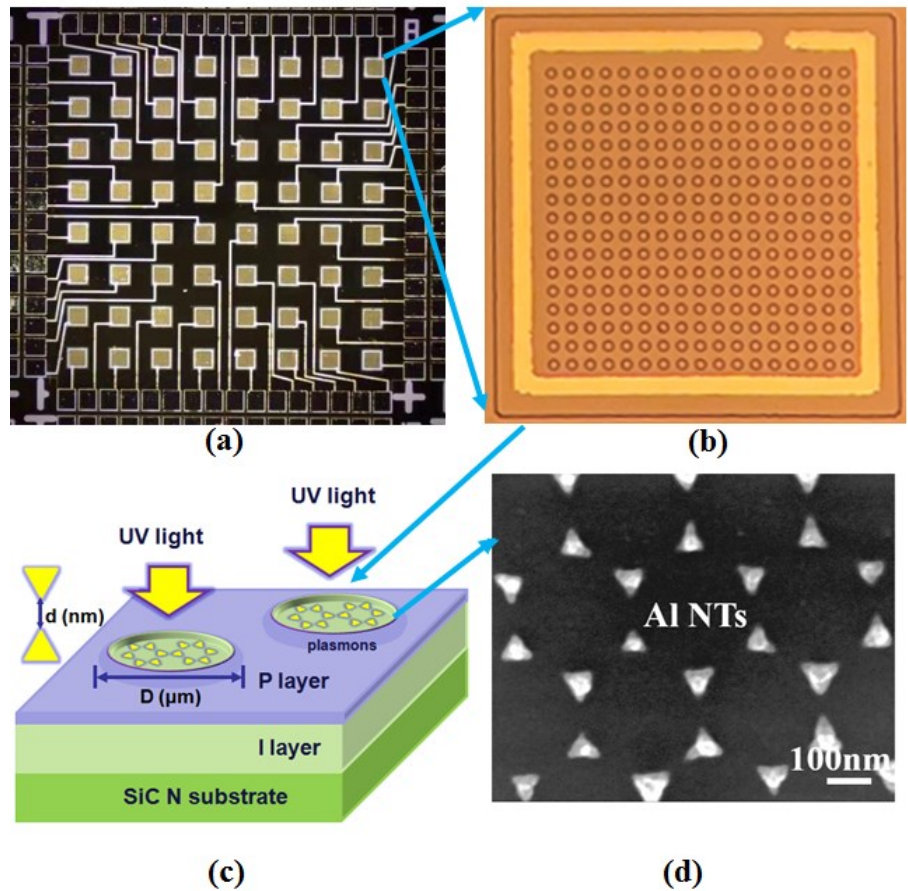

**Fig. 1 (a) Top-view of the whole 8×8 p-i-n APD array with MHs and Al NTs; (b) Single p-i-n APD pixel with MHs and Al NTs; (c) Magnified schematic of 4H-SiC p-i-n APD with MHs and Al NTs; (d) Section morphology and magnified view of Al NTs arrays in the MHs.**

**Design and Experiment**

The 4H-SiC avalanche photodetectors (APDs) with micro-hole (MH) structures and aluminium nano-triangles (NTs) were fabricated into an 8×8 array with dimensions of 2×2 mm², as illustrated in Fig. 1(a). The effective photosensitive area of a single pixel was designed to be 200×200 μm², as shown in Fig. 1(b). The top surface of each device was etched with the MH structure, which serves as a window to enhance ultraviolet (UV) absorption. The MHs were etched through to the intrinsic i layer with diameters (D) of 4 μm, 8 μm, and 10 μm, respectively. As depicted schematically in Fig. 1(c), the APD structure was processed using standard 4H-SiC epitaxial techniques. It consists of a 200 nm thick p-type layer ($N_A=1\times10^{19}$ cm$^{-3}$), a 3000 nm thick i layer ($N_D=1\times10^{15}$ cm$^{-3}$) on a 365 μm thick n substrate ($N_D = 5 \times 10^{18}$ cm$^{-3}$). Notably, in contrast to conventional APD structures, the i layer of the photodiode in this design is directly exposed to UV light, enabling enhanced UV detection capabilities. The proposed APD design incorporates an array of Al hexagonal NTs within the MHs, positioned on the 4H-SiC i-layer, as shown in Fig. 1(d). The diameter (D) of the MHs and the spacing (d) between the NTs were varied from 4 μm to 10 μm and 10 nm to 80 nm, respectively. The Al NTs were designed with a side length of 80 nm and a thickness of 10 nm. The geometric features of the Al NTs could be precisely controlled using a polystyrene (PS) microsphere template[42,43]. Further details regarding the fabrication process are provided in the Methods section of the Supplementary Information.

## Results and Discussions

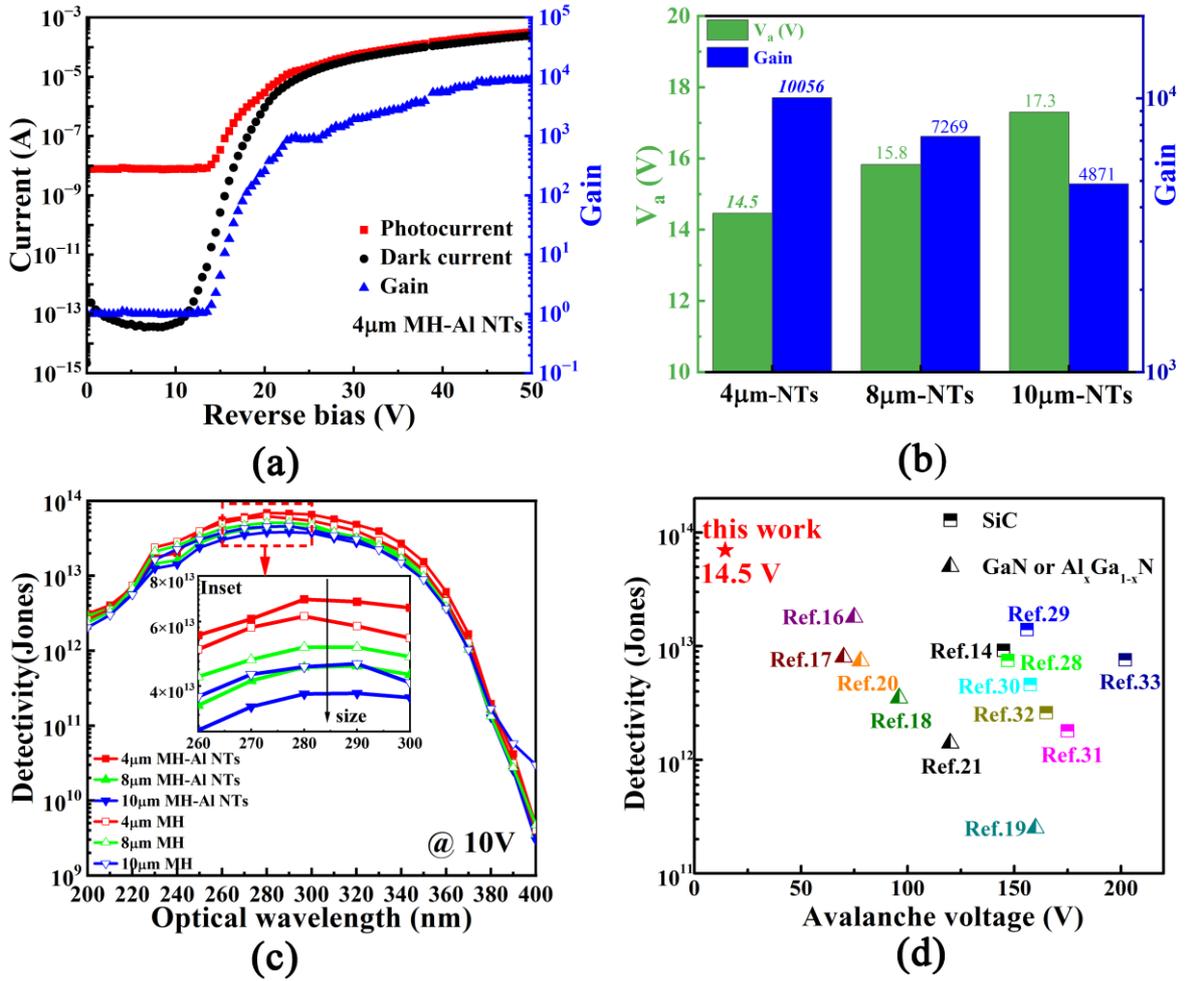

Fig. 2 (a) The photocurrent, dark current and gain versus voltage characteristics of the 4H-SiC p-i-n APDs with 4 μm MH and Al NTs. (b) The statistical comparison histograms of avalanche breakdown voltage ($V_a$) and gain for the APDs with Al NTs for different diameters MHs (4 μm, 8 μm and 10 μm). (c)The corresponding detectivity with the wavelength range from 200 nm to 400 nm for the APDs with and without Al NTs at 10 V reverse bias. The inset shows detectivity from 260 nm to 300 nm. (d) Comparison of detectivity at unity gain and avalanche breakdown voltage with other work (including 4H-SiC, GaN and $Al_xGa_{1-x}N$ APDs).

Current-voltage (I-V) characteristics were measured to assess the performance of the proposed APDs. These measurements revealed that the local avalanche effect occurs at a reverse bias of approximately 14.5 V in devices with aluminum NTs in the SiC MHs, as depicted in Fig. 2(a). The reduction in avalanche voltage offers significant advantages, particularly in preventing instantaneous breakdown and safeguarding the device from catastrophic damage. In addition, the photocurrent in devices incorporating Al NTs in the MHs was observed to be within the range of $10^{-5} \sim 10^{-3}$ A for reverse biases between 20 V and

50 V, compared to devices without Al NTs, which exhibited photocurrents of approximately $10^{-9} \sim 10^{-8}$ A. The gain of devices with Al NTs reached $10^4$ within a voltage range of a few tens of volts, whereas the gain in devices without Al NTs was significantly lower, only increasing by 6 times (Figs. 2(a) and S2). Optimization of device performance was achieved by adjusting the MH diameters, with the 4 μm MH devices incorporating Al NTs exhibiting the lowest local avalanche starting voltage of 14.5 V and the highest gain of 10,056 (Fig. 2(b)). Detailed plots of the photocurrent, gain, and dark current for the Al NT devices are provided in Figs. S1–S3.

Furthermore, the detectivity of the devices was evaluated across various optical wavelengths (Fig. 2(c)). Devices with 4 μm MHs and Al NTs achieved the highest detectivity of $7.0 \times 10^{13}$ Jones at a wavelength of 280 nm under a reverse bias of 10 V. Notably, the detectivity decreased as the MH size increased, regardless of the presence of Al NTs. This reduction in detectivity can be attributed to changes in the depletion region, which negatively impacts the collection of carriers.

A comparison of detectivity versus avalanche voltage was conducted to benchmark our devices against existing wide-bandgap semiconductor technology (Fig. 2(d)). Our APDs demonstrated superior performance, featuring the highest detectivity and the lowest avalanche voltage of 14.5 V. These characteristics underscore the exceptional performance metrics, including low avalanche voltage, high detectivity, and high gain. This balance is particularly advantageous in the realm of micro- and nano-photonics, where surface plasmon excitations in MHs are leveraged in 4H-SiC APDs. Moreover, the stability of the device was evaluated to determine the impact of long-term operation on performance. After 13,000 hours (1.5 years) of exposure, multiple I-V measurements showed consistency with the initial curve (Fig. S4). These results indicate that the proposed devices, incorporating Al NTs, exhibit robust and stable operation in the local avalanche state without the need for quenching circuits.

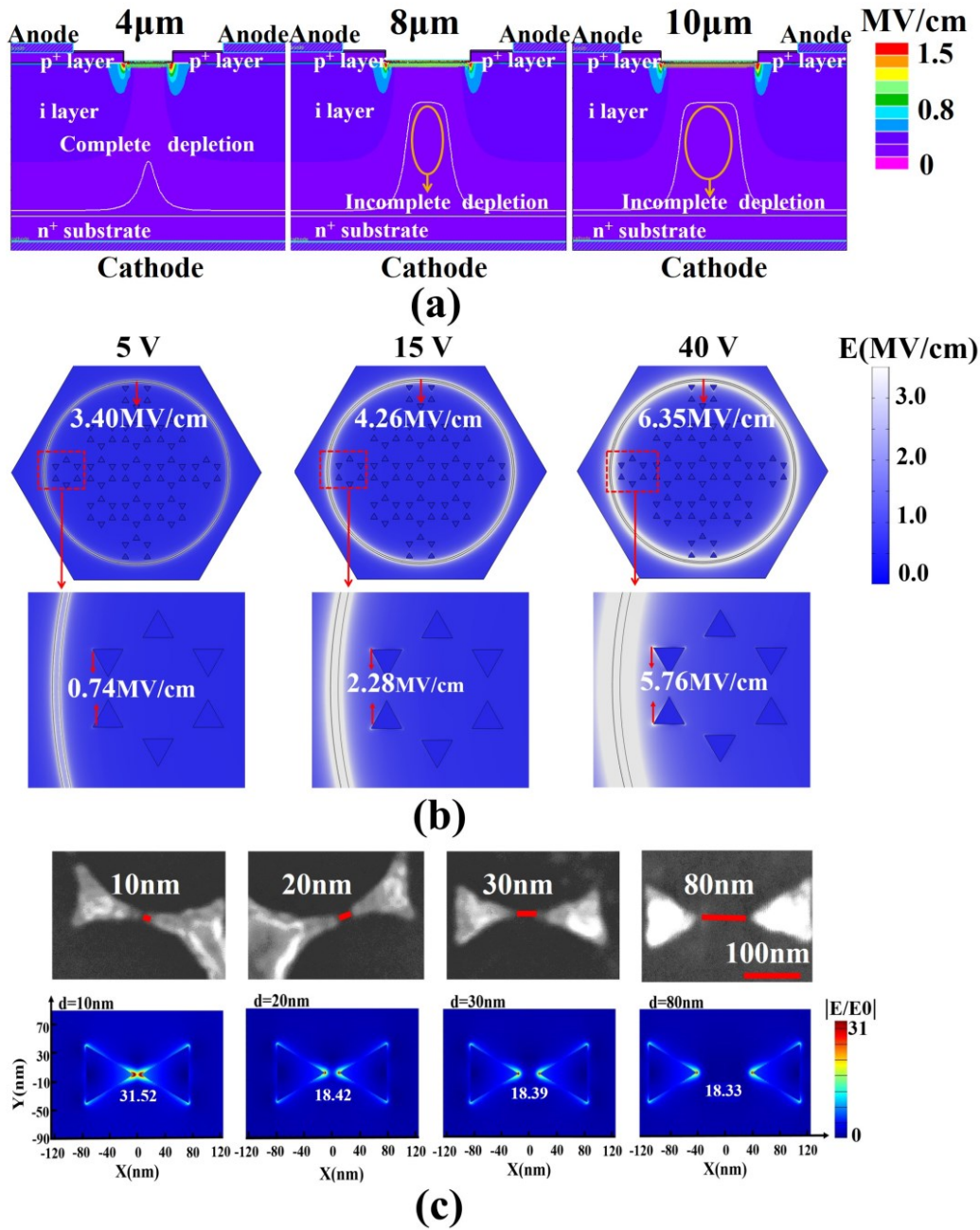

**Fig.3 (a)** The internal electric field intensity of the devices with MH (4 μm, 8 μm and 10 μm) and Al NTs under the charge accumulation effect at the tips of Al NTs was simulated by applying 15 V reverse bias. **(b)** The electric field intensity at the tips of Al NTs was simulated at different reverse bias (5 V, 15 V and 40 V). **(c)** FDTD simulations of field enhancement of Al NTs under illumination with different distance corresponding to SEM images of NTs with distance of 10 nm, 20 nm, 30 nm, and 80 nm.

The electric field is a crucial factor in determining carrier collection and multiplication within APDs. Therefore, a theoretical simulation based on finite element analysis (FEA) was conducted using TCAD to investigate the electric field distribution in the absence of illumination (Figs. 3(a) and 3(b)). The bulk electric field distribution for the APDs with different MH sizes (Fig. 3(a)) indicates that a complete depletion layer is formed at the junctions for devices with 4 μm MHs and Al NTs. In contrast, the intrinsic i layer is only partially depleted in APDs with 8 μm and 10 μm MHs and Al NTs. The undepleted regions serve as recombination centers, resulting in reduced carrier collection during avalanche events, which explains why the APDs with 4 μm MHs and Al NTs exhibit superior performance in terms of gain, detectivity, and avalanche voltage (Figs. 2(b) and 2(c)).

The local field enhancement induced by the lightning-rod effect near the corner of MH becomes prominent under different applied voltages. Under dark conditions, the local field enhancement at the edges of the MH, induced by the lightning-rod effect, becomes substantial at different applied voltages. For instance, at a reverse bias of 5 V, electric field aggregation is observed at the tips of the aluminium nanotubes (Al NTs) near the electrode. However, the electric field intensity reaches only 0.74 MV/cm (Fig. 3(b)), which remains insufficient to initiate a local avalanche effect within the device. As the reverse bias increases to 15 V, the electric field intensity at the tips of the Al NTs rises to 2.28 MV/cm, causing the internal electric field of the device to reach 1.5 MV/cm (Fig. 3(b)), which is sufficient to induce a local avalanche effect. With further increases in the reverse bias, the electric field intensity at the tips of the Al NTs continues to strengthen; at 40 V, it reaches a high value of 5.76 MV/cm, which leads to an expanded avalanche region and an increase in avalanche gain. Invariably, the regions where local avalanches occur remain concentrated at the edges of the MHs, as the electric field intensity at the Al NT tips diminishes with distance from the electrodes.

In contrast, for devices without Al NTs, the surface electric field intensity is approximately one-fifth of that observed in devices containing Al NTs. At a 15 V reverse bias, the surface electric field intensity at the edges of the MHs is only 0.41 MV/cm (Fig. S5), with the corresponding internal electric field within the device remaining below 0.4 MV/cm (Fig. S6), which is not adequate to trigger a local avalanche and yield higher gain. These field enhancement effects at the nanotube tips have been corroborated by other studies[41].

To further explore the local field enhancement mechanism of Al NTs under illumination, we employed the finite difference time domain (FDTD) method to simulate and validate the field intensity distribution of Al NTs. The spacing between Al NTs was set at 10 nm, 20 nm, 30 nm, and 80 nm, with the side length and thickness of the NTs set at 80 nm and 10 nm, respectively, in alignment with the experimental parameters. It was found that the spacing between adjacent NTs strongly influences the field enhancement (Fig. 3(c)). Specifically, when the spacing is reduced to less than 10 nm, the plasmon coupling field intensity increases by more than 30-fold. The strong confinement of the enhanced impact ionization at smaller gaps between NTs further extends the local avalanche region, enhancing performance of the device at the local avalanche state. Although plasmonic coupling between bow-tie structures diminishes as the NT spacing increases from 20 nm to 80 nm, the field intensity remains more than 18 times greater than that of untreated tips. This field intensity is sufficient to cooperate avalanche multiplication, confirming the effectiveness of the Al NTs surface plasmon excitations in coupling the high local electric field and improving device performance.

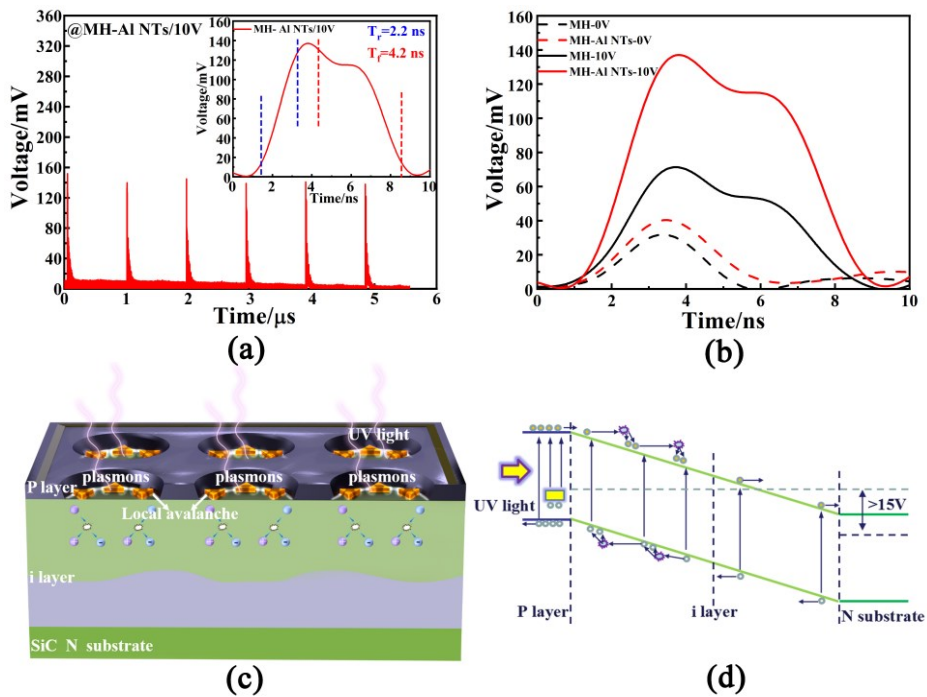

**Fig.4 (a) The response speed of the 4 µm-MH devices with Al NTs at 10 V and the inset shows the single impulse response of the device with Al NTs at 10 V. (b) Comparisons of single impulse of the devices with and without Al NTs at 0 V and 10 V. (c) Local avalanche schematics and (d) band diagram of the APD with MHs and Al NTs under the condition of illumination.**

The temporal spectral response of the 4H-SiC APDs was investigated using a 266 nm picosecond laser as the illumination source and an oscilloscope for detection. The rise time ($T_r$) and fall time ($T_f$) were defined as the times during which the impulse voltage rises from 10% to 90% and falls from 90% to 10% of the peak value, respectively. Figure 4(a) and its inset present the photocurrent response at an applied reverse bias of -10 V. The results demonstrate that the APD with Al NTs exhibits superior operational stability and reliability. Additionally, due to the p-i-n vertical structure and the inherently low minority carrier lifetime of the 4H-SiC material, the devices incorporating 4 μm MHs and Al NTs achieved a high response speed, with a $T_r$ of 2.2 ns and a $T_f$ of 4.2 ns (Fig. 4(a) inset). The $T_r$ of the proposed device represents an 82.2% reduction compared to the 12.4 ns response time reported for $Ga_2O_3$-based photodetectors[23]. Furthermore, it was observed that both devices, with and without Al NTs, exhibited longer response times at a reverse bias of 10 V compared to 0 V (Fig. 4(b) and Fig. S7). Specifically, the $T_f$ of 4.2 ns at 10 V reverse bias was more than twice the 1.9 ns observed at 0 V. This increase is attributed to the acceleration of photogenerated carriers within the space-charge region at higher reverse bias, necessitating a longer response time. Moreover, both $T_r$ and $T_f$ in devices with Al NTs increased by approximately 10% compared to devices without Al NTs at a reverse bias of -10 V. These temporal performance enhancements highlight the strong carrier multiplication induced by the local field enhancement effect and demonstrate voltage-tunable response times in the proposed Al NT-integrated device. The combination of fast temporal response and low operational voltage positions the device as a promising candidate for high-speed ultraviolet (UV) detection applications.

The local avalanche mechanism and carrier transport in UV-illuminated devices are further elucidated through the schematic and band diagrams (Figs. 4(c) and 4(d)). The upper corner region of the device exhibits higher electric field, corresponding to the local avalanche area. In this region, photogenerated carriers undergo collision ionization, resulting in the generation of electron-hole pairs. These pairs are efficiently separated by the reverse-biased electric field of pn junction. In contrast to conventional APDs, the collision ionization process in the proposed device is localized to a specific area, while the remainder of the device remains stable. As a result, the photocurrent tends to saturate with increasing reverse bias, indicating that the device operates in a local avalanche state rather than a full avalanche state.

**Conclusion and outlook**

This study effectively enhanced the local electric field of 4H-SiC avalanche photodiodes featuring MHs by integrating the lightning rod effect with localized surface plasmon resonance of Al NTs. This approach achieved an unprecedentedly low avalanche initiation voltage of 14.5 V. Notably, these devices can operate in a local avalanche state without requiring quenching circuits, maintaining a high avalanche gain of $10^4$ and demonstrating excellent long-term stability.

The concept of local avalanche within 4H-SiC APDs introduces an innovative method for internal electric field regulation in semiconductor devices and holds potential for adaptation to other photodetectors based on wide-bandgap semiconductors. Future directions include optimizing the shape of metal nanostructures, with an emphasis on sharper features and optimised curvature radii, to maximize both the lightning rod effect and surface plasmon resonance, thereby further improving device performance. These findings indicate that 4H-SiC materials and associated technologies constitute a robust and scalable platform for quantum-based light detection and imaging, offering ease of integration and high compatibility with industrial applications.

**Acknowledgement**

This work was supported by the National Key Research and Development Program of China (Grant No.2018YFB0905700), the State Key Laboratory of Advanced Power Transmission Technology (Grant No. GEIRI-SKL-2022-005), National Natural Science Foundation of China (Grant No. 62274137 and 62104222), Natural Science Foundation of Jiangxi Province of China for Distinguished Young Scholars (No. S2021QNZD2L0013), the Fundamental Research Funds for the Central Universities (Grant No. 20720230103,Grant No. 20720220026), Jiangxi Provincial Natural Science Foundation (Grant No. 20232BAB202043), the Science and Technology Project of Fujian Province of China (Grant No. 2020I0001).